\documentclass[10pt,conference]{IEEEtran}
\IEEEoverridecommandlockouts
\usepackage{cite}
\usepackage{amsmath,amssymb,amsfonts}
\usepackage{algorithmic}
\usepackage{graphicx}
\usepackage{textcomp}
\usepackage{xcolor}
\def\BibTeX{{\rm B\kern-.05em{\sc i\kern-.025em b}\kern-.08em
    T\kern-.1667em\lower.7ex\hbox{E}\kern-.125emX}}
    
    \newcommand{\sectopic}[1]{\vspace{0.2em}\par\noindent{\textit{\bfseries #1}}}
    
\begin{document}

\title{PersonaGen: A Tool for Generating Personas from User Feedback}

\author{
    \IEEEauthorblockN{
    Xishuo Zhang\IEEEauthorrefmark{1}, 
    Lin Liu\IEEEauthorrefmark{1}, 
    Yi Wang\IEEEauthorrefmark{2}, 
    Xiao Liu\IEEEauthorrefmark{2},
    Hailong Wang\IEEEauthorrefmark{1},
    Anqi Ren\IEEEauthorrefmark{1},
    Chetan Arora\IEEEauthorrefmark{3}
}

   \IEEEauthorblockA{\IEEEauthorrefmark{1}College of Computer Science and Technology, Inner Mongolia Normal University, Hohhot, China
    \\ xishuozhang163@163.com, liulin@imnu.edu.cn, 9177385@qq.com, 3138363067@qq.com}
    \IEEEauthorblockA{\IEEEauthorrefmark{2}School of Information Technology, Deakin University, Geelong, Australia
    \\ xve@deakin.edu.au, xiao.liu@deakin.edu.au}
    \IEEEauthorblockA{\IEEEauthorrefmark{3}Faculty of Information Technology, Monash University, Melbourne, Australia
    \\ chetan.arora@monash.edu}
   
}

\maketitle

\begin{abstract}
Personas are crucial in software development processes, particularly in agile settings. However, no effective tools are available for generating personas from user feedback in agile software development processes. To fill this gap, we propose a novel tool that uses the GPT-4 model and knowledge graph to generate persona templates from well-processed user feedback, facilitating requirement analysis in agile software development processes. We developed a tool called PersonaGen. We evaluated PersonaGen using qualitative feedback from a small-scale user study involving student software projects. The results were mixed, highlighting challenges in persona-based educational practice and addressing non-functional requirements.

\end{abstract}

\begin{IEEEkeywords}
 Persona, GPT-4 Model, Knowledge Graph, Requirements Engineering, User Feedback.
\end{IEEEkeywords}

\section{Introduction}\label{sec:introdcution}

Personas are critical in software development, particularly in agile settings, to gather information about user requirements, resolve potential requirements issues and foster the coverage of diverse and inclusive requirements~\cite{karolita2023use}. Among various data sources for creating personas, user feedback is indispensable and can be collected through, among others, app reviews, interviews, surveys, and usability testing~\cite{karolita2023use}. However, using personas in agile software development based on user feedback has several challenges. For instance, development teams must consistently update and maintain these personas when requirements change~\cite{arora2015change}. 
Moreover, in the context of rapid iteration and delivery in agile development, using personas can undeniably be seen as a waste of time, especially for startups. Thus, it is pivotal to effectively generate personas from user feedback data in agile software development processes.

Personas and user feedback jointly contribute to user experience (UX), user interface design (UI), and software development processes \cite{Joni22}. Notably, user feedback can be employed to optimize personas, and combining personas and user feedback can facilitate the iteration of software projects. In recent years, some research has started focusing on applying data-driven personas in various types of software projects, such as B2B (Business to Business) software projects \cite{Watanabe18}. However, there is still a gap in exploring the use of feedback data for persona generation in agile software development processes. Therefore, using user feedback to generate personas is an imminent need in agile software development processes.

To address this research gap, we propose a tool called PersonaGen. This tool employs the GPT-4 model and knowledge graphs for generating personas from well-processed user feedback. PersonaGen has three major features. \textbf{\textit{1.}} PersonaGen is the first tool to use the GPT-4 model for cleaning, integrating, predicting and analyzing user feedback. \textbf{\textit{2.}} PersonaGen is the first to construct a knowledge graph through various data attributes. \textbf{\textit{3.}} PersonaGen can classify different persona attributes and establish connections between various attributes to generate persona recommendations.




\section{PersonaGen Description}\label{sec:Architecture}

\begin{figure}[!t]
  \centering
  \includegraphics[width=\linewidth]{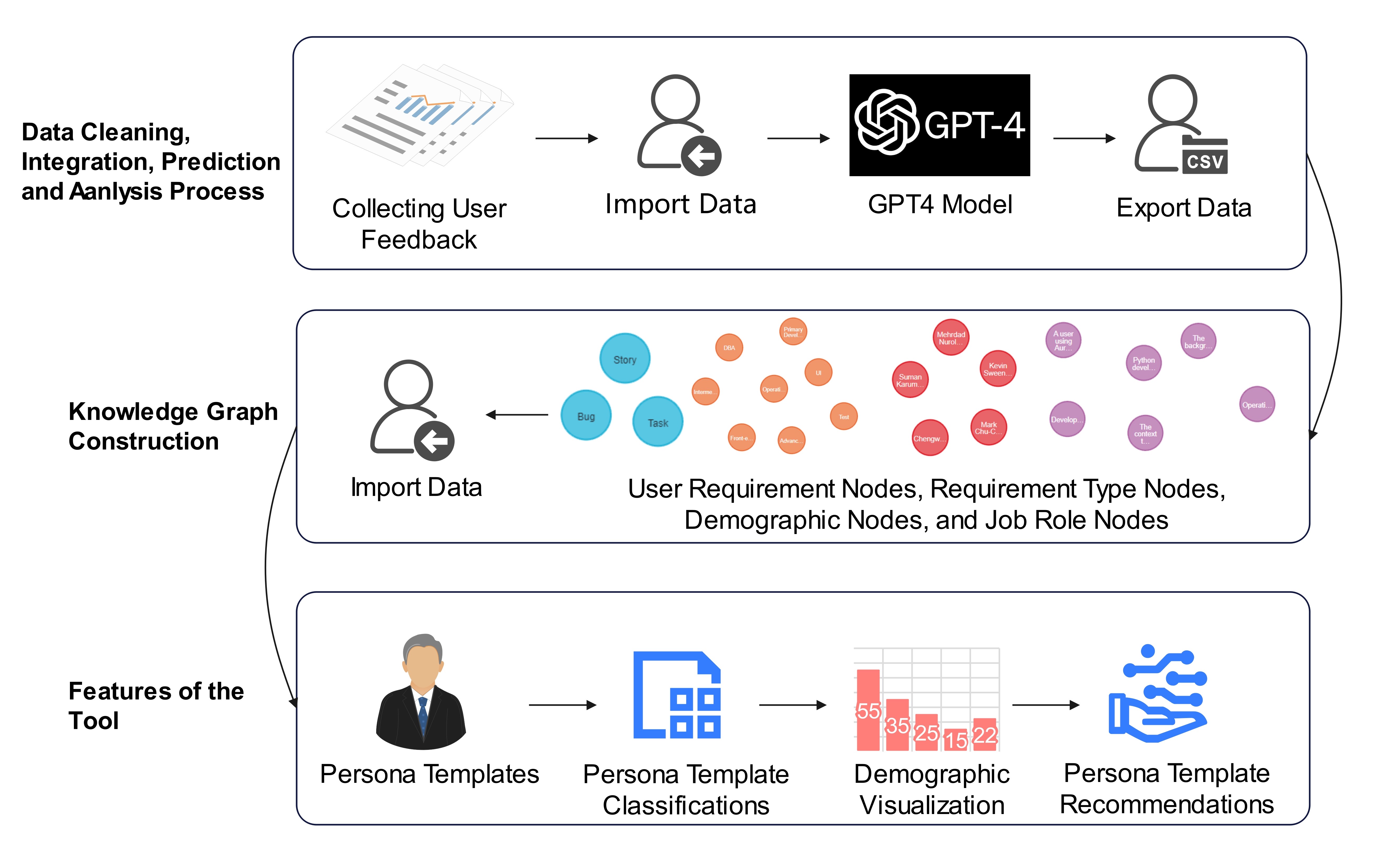}
  \vspace*{-1em}
  \caption{The Main Processes of the PersonaGen Tool. }  
   \label{fig:personat}
   \vspace*{-1em}
\end{figure}

The main objective of PersonaGen is to help agile development teams to generate personas from user feedback. Personas are a crucial tool in agile software development processes. Personas can help teams in understanding potential user needs and defining the software features. Our goal is to use the GPT-4 model and construct a knowledge graph to develop a tool capable of generating persona templates from well-processed data. Fig.~\ref{fig:personat} illustrates the main processes of PersonaGen.



\subsection{Major Features}\label{subsec:features}

\sectopic{1. User Feedback Data Cleaning, Integration, Prediction, and Analysis via GPT-4 Model. }An important process in our tool involves cleaning, integration, predicting, and analyzing user feedback. Effectively processing is a key step in generating high-quality persona templates. To facilitate this, we employed the GPT-4 model into our tool, enabling the generation of various high-quality and detailed persona contents. The tool can save the processed data in CSV format.


\sectopic{2. Knowledge Graph Construction-Based User Feedback Data. }We constructed a knowledge graph to strengthen the connections among various attributes associated with each persona. We built different nodes with various data attributes, to facilitate persona classifications and recommendations. These include user requirement nodes, requirement type nodes, demographic nodes, and job role nodes. Furthermore, users can define different nodes based on their specific preferences. 

\begin{figure}[!t]
  \centering
  \includegraphics[width=\linewidth]{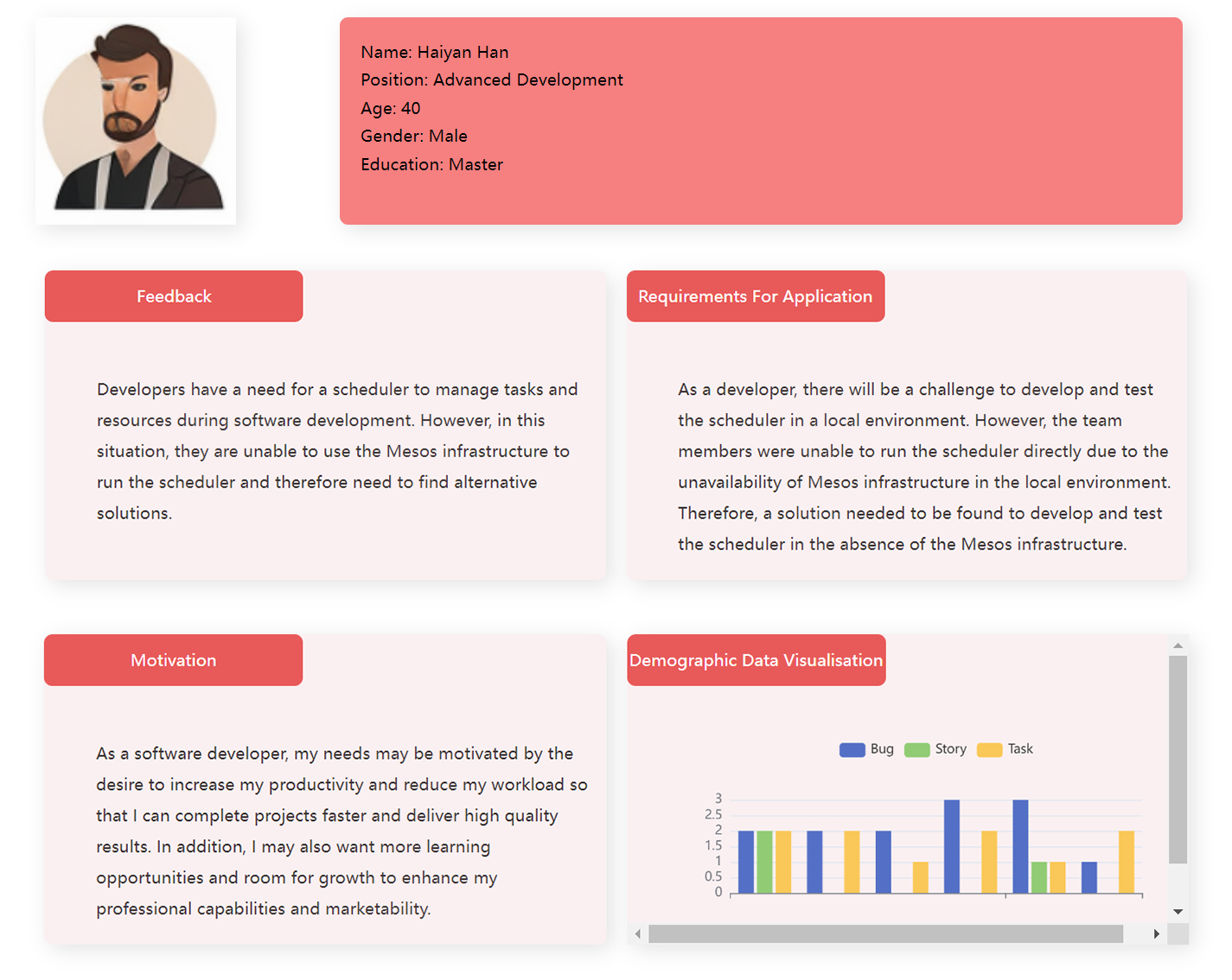}
  \vspace*{-0.5em}
  \caption{An Example of Persona Template. }
   \label{fig:example}
   \vspace*{-0.5em}
\end{figure}

\sectopic{3. Generating Persona Templates. }We developed persona classifications and recommendations. Our tool can classify different persona templates and recommend similar personas based on various attributes. Moreover, the content of these personas includes demographic information, visualization, feedback, motivations and requirement for applications from user feedback. Fig.~\ref{fig:example} illustrates an example persona template.




\subsection{Implementation}

We developed PersonaGen as a web application. PersonaGen has been implemented using common web technologies (HTML, CSS, JavaScript, and Java-Springboot). The database for constructing the knowledge graph uses Neo4j\footnote{https://neo4j.com}. We first applied the GPT-4 model\footnote{https://openai.com/gpt-4} for cleaning, integrating, predicting, and analyzing user feedback obtained from student software projects. The tool project is publicly available\footnote{https://github.com/xishuozhang/PersonaGen/tree/main}.

\section{Evaluation Results}\label{sec:evaluation}

We conducted small-scale user studies to evaluate the value of PersonaGen. Specifically, we analyzed the qualitative feedback data from three student software projects, which involved a total of 13 third-year of undergraduate participants/students. We generated a series of persona templates using PersonaGen based on user feedback from student software projects. We found that the results were rather mixed. Some participants considered that the accuracy of persona generation using the GPT-4 model was superior to an independent analysis conducted by the participants themselves, primarily due to their lack of rich experience in qualitative analysis. In addition, some participants had some experience in software development within the industry and had a better understanding of the concept and application of personas. However, some participants considered that there was a lack of education and practical knowledge related to persona-based practices in student projects. While most participants were confident in effectively addressing functional requirements, they found it challenging to analyze non-functional requirements (NFRs), such as users with accessibility requirements.


\section{Conclusion and Future Work}\label{sec:conclusion}

This paper introduces a tool called \textit{PersonaGen} to assist agile software development processes in using personas for RE. The core module of this tool relies heavily on the large language model provided by the GPT-4 model. The construction of a knowledge graph facilitates the classification and recommendation of persona templates based on different nodes. The feedback on the PersonaGen tool was rather mixed, suggesting a need for enhanced persona-based educational practices, and highlighting challenges in addressing NFRs. Furthermore, we plan to focus on addressing accessibility requirements and integrating more human-centric aspects into persona templates. In the future, this PersonaGen tool will be employed in educational practices, such as UI/UX design and requirements engineering-related courses.

\section*{Acknowledgements}

This work is supported by the Natural Science Foundation of Inner Mongolia (Grand No.2023LHMS060062020), National Key Research and Development Plan (Grand No.2020YFC1523305).

\bibliographystyle{IEEEtran}

\bibliography{ref}

\clearpage

\end{document}